\documentstyle[prbbib,aps,twocolumn]{revtex}
\input{psfig}
\begin{document}
\topmargin -1.cm
\baselineskip 0.3in 
\centerline{\large {\bf A Microscopic Model of Gemini Surfactants:}}
\centerline{\large {\bf Self-assemblies in Water and at Air-Water Interface}} 

\vspace{.25cm} 

\centerline{\bf Prabal K. Maiti$^{1}$} 
\centerline{Department of Physics, Indian Institute of Technology,}
\centerline{ Kanpur 208016, U.P., India}

\vspace{.15cm}

\centerline{and} 

\vspace{.15cm}

\centerline{\bf Debashish Chowdhury$^{2,*}$}
\centerline{Institute for Theoretical Physics, University of Cologne,}
\centerline{ D-50937 K\"oln, Germany}
\centerline{and}
\centerline{Department of Physics, Indian Institute of Technology,}
\centerline{ Kanpur 208016, U.P., India$^{\dagger}$}
\begin{abstract}
We report the results of large scale Monte Carlo (MC) simulations of 
novel microscopic models of gemini surfactants to elucidate (i) their  
spontaneous aggregation in bulk water and (ii) their spatial organization 
in a system where water is separated from the air above it by a sharp 
well defined interface. We study the variation of the critical micellar 
concentration (CMC) with the variation of the (a) length of the spacer, 
(b) length of the hydrophobic tail and (c) the bending rigidity of the 
hydrocarbon chains forming the spacer and the tail; some of the trends 
of variation are counter-intuitive but are in excellent agreement with 
the available experimental results. 
Our simulations elucidate the effects of the geometrical shape, size 
and density of the surfactant molecules, the ionic nature of the 
heads and hydrophobicity/hydrophilicity of the spacer not only on the 
shapes of the micellar aggregates and the magnitude of the CMC, but also   
on their conformations close to the air-water interface. 
\end{abstract}
{\bf Running title:} Self-assemblies of gemini surfactants 

------------------------------------------------

$^{1}$ E-mail: prabal@iitk.ernet.in 

$^{2}$ E-mail: debch@iitk.ernet.in 

$^{*}$ To whom all correspondence should be addressed. 

$^{\dagger}$ Present and permanent address.

\vfill\eject 

\section{Introduction:} 

Soap molecules are common examples of surfactant molecules; these not
only find wide ranging applications in detergent and pharmaceutical 
industries, food technology, petroleum recovery etc. but are also one 
of the most important constituents of cells in living systems. Therefore, 
physics, chemistry, biology and technology meet at the frontier area of 
interdisciplinary research on association colloids formed by surfactants 
\cite{Evans}. 

\begin{figure}
\centerline{\psfig{file=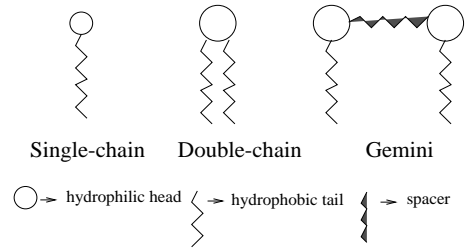,width=6cm}}
\caption{\sf Different types of amphiphiles.}
\label{fig1a}
\end{figure}

The "head" part of surfactant molecules consist of a polar or ionic group. 
The "tail" of many surfactants consist of a single hydrocarbon chain 
whereas that of some other surfactants, e.g., phospholipids, are made 
of two hydrocarbon chains both of which are connected to the same head 
~\cite{tanford}. In contrast, {\it gemini} surfactants 
~\cite{Deinega,Menger1,Menger2,Rosen1}, 
consist of two single-chain surfactants whose 
heads are connected by a "spacer" chain and, hence, these "double-headed" 
surfactants are sometimes also referred to as "dimeric surfactants" 
\cite{Zana1,Zana2} (see fig.1(a)). 
When put into an aqueous medium, the "heads" of the surfactants like to get 
immersed in water and, hence, are called "hydrophilic" while the tails tend to 
minimize contact with water and, hence, are called "hydrophobic" \cite
{tanford}. The spacer in gemini surfactants is usually hydrophobic but gemini 
surfactants with hydrophilic spacers have also been synthesized\cite{Rosen2}. 
Surfactant molecules are called "amphiphilic" because their heads are 
"water-loving" and hydrocarbon chains are "water-hating".
Because of their amphiphilicity the surfactant molecules form "{\it 
self-assemblies}"  (i.e., supra-molecular aggregates), such as monolayer 
and bilayer membranes, micelles, inverted-micelles, etc. \cite{Gelbert},  
in a multi-component fluid mixture containing water. 

In this paper we develope a microscopic model of {\it gemini} surfactants 
and, carrying out Monte Carlo (MC) computer simulations, investigate how 
the shapes and sizes of these molecules as well as their mutual interactions 
and their interactions with the molecules of water give rise to some  
unusual aggregation phenomena. Another aim of this paper  
is to report the results of a complimentary MC study of the spatial 
organization of these model gemini surfactants  (particularly, their tails 
and spacers) at the air-water interface in order to answer some of the 
fundamental questions raised on this point and speculations made in the 
literature. 

Micelles are formed when the concentration of the surfactants in water 
exceeds what is known as the critical micellar concentration (CMC) 
\cite{tanford}. In reality, CMC is not a single concentration  
(it is more appropriate to call it characteristic micellar concentration
~\cite{St1}). A longer hydrocarbon chain leads to larger area of contact 
between water and the hydrophobic part of an isolated surfactant molecule.
Therefore, intuitively, one would expect that a longer hydrocarbon chain 
should enhance the tendency for aggregation, i.e., lower the CMC. This 
is, indeed, in agreement with the trend of variation of CMC of single-chain 
surfactants with the length of the hydrophobic tail~\cite{Evans}. 
In contrast, two unusual features of the CMC of gemini surfactants with 
{\it ionic heads} and {\it hydrophobic spacer} are:\\ 
(i) for a given fixed length of each of the two tails, the CMC 
{\it increases} with the length of the spacer till it reaches a maximum 
beyond which CMC {\it decreases} with further increase of the spacer length  
\cite{Zana1,Zana3,Frindi,goyel};\\
(ii) for a given length of the spacer, the CMC {\it increases} with 
increasing tail length \cite{Menger1,Menger2}.\\ 
Moreover, the micellar aggregates formed by the gemini surfactants with 
short spacers even at low concentrations just above the CMC are "long, 
thread-like and entangled" \cite{Zana2,Zana4}, in contrast to the spherical 
shapes of the micelles formed by single-chain surfactants at such low 
concentrations. Furthermore, the CMC of gemini surfactants with {\it ionic 
heads} and {\it hydrophilic spacer} decrease monotonically with the increase 
of the length of the spacer~\cite{Zhu}. Our aim is to understand the 
physical origin of these unusual properties of gemini surfactants. 
We also make some new predictions on the morphology of the micellar 
aggregates of gemini surfactants with long tails and long spacer. 

Synthesis of gemini surfactants with non-ionic (polar) heads in laboratory 
experiments remains one of the challenging open problems. But, our 
computer experiments on model gemini surfactants with non-ionic heads 
enable us to predict their morphologies and the variation of their CMC 
with the lengths of tails and spacers. 

In the presence of the air-water interface, where do the tails and the 
spacer of an isolated gemini surfactant find themselves- do they lie inside 
water or outside (i.e., in the air), do they get crowded close to the 
interface or do they spread out as far away from the interface as 
possible? How does the effective area of cross-section of an isolated 
gemini surfactant at the air-water interface vary with the increase of the 
length of the spacer when the spacer is (a) hydrophobic, (b) hydrophilic? 
How do the conformations of the gemini surfactants and spatial 
organizations of their tails and spacers vary with the increase of the 
density which gives rise to unavoidable interactions (both direct and 
entropic) among the surfactants. We try to answer these fundamental 
questions by carrying out computer experiments on a microscopic model that 
we propose here. 

In this paper we simulate gemini surfactants with (a) hydrophobic spacers 
and also those with (b) hydrophilic spacers. For each of these two types 
of gemini surfactants we consider both ionic and non-ionic (but polar) 
hydrophilic heads. 

A microscopic lattice model of double-chain surfactants (with a single 
head) in aqueous solution was developed by Bernardes\cite{bern2} by 
modifying the Larson model of single-chain surfactants~\cite{Lar1,St1,Livrev}. 
In this paper we propose a microscopic lattice model of gemini surfactants 
by extending Bernardes' model so as to incorporate two hydrophilic heads  
connected by a {\it hydrophobic spacer}. A summary of the important 
preliminary results for the gemini surfactants with {\it hydrophobic} 
spacers has been reported elsewhere. In this paper we not only give details 
but also report the corresponding results for gemini surfactants with 
{\it hydrophilic} spacers and study the effects of a third component 
(an oil) on the novel morphology of the micellar aggregates formed by the 
gemini surfactants in water. 

A microscopic model for single-chain surfactants at the air-water 
interface was developed earlier by one of us \cite{Chow1,Chow2} by 
appropriately modifying the Larson model \cite{Lar1,St1,Livrev} of 
ternary microemulsions \cite{Chen,Gelbert}. In this paper we replace the 
single-chain surfactants in the model introduced in ref.\cite{Chow1} by 
the model gemini surfactants developed here, thereby getting the desired 
microscopic model of gemini surfactants at the air-water interface.

The model and the characteristic quantities of interest are defined in 
section 2. The results on the micellar aggregates, in bulk water, formed 
by the gemini surfactants with hydrophobic and hydrophilic spacers are 
reported in two subsections of section 3. The results of the investigations 
on the spatial organization of the gemini surfactants with hydrophobic and 
hydrophilic spacers at the air-water interface are given in section 4. 
Finally, a summary of the main results and the conclusions drawn from these 
are given in section 5.  

\section{The Model and The Characteristic Quantities of Interest:}

\subsection{General aspects}

The Larson model was originally developed for ternary microemulsions which 
consist of water, oil and surfactants. In the spirit of lattice gas models, 
the fluid under investigation is modelled as a simple cubic lattice of size 
$L_x \times L_y \times L_z$. Each of the molecules of water (and oil) can  
occupy a single lattice site. A surfactant occupies several lattice sites 
each successive pairs of which are connected by a rigid nearest-neighbour bond. 
A single-chain surfactant can be described by the symbol ~\cite{Livrev}
${\cal T}_m{\cal N}_p{\cal H}_q$ where ${\cal T}$ denotes {\it tail}, 
${\cal H}$ denotes {\it head} and ${\cal N}$ denotes the 'liaison' or neutral 
part of the surfactants. $m$, $p$ and $q$ are integers denoting the lengths 
of the tail, neutral region and head, respectively, in the units of lattice 
sites. Thus, each single-chain surfactant is a self-avoiding chain of 
length $\ell = (m+p+q)$. We shall refer to each site on the surfactants as a 
$monomer$. The "water-loving" head group is assumed to be "water-like" 
and, similarly, the "oil-loving" tail group is assumed to be "oil-like". 

Jan, Stauffer and collaborators \cite{St1} simplified the Larson model 
by formulating it in terms of Ising-like variables which interact with 
nearest-neighbour interaction $J$, in the same spirit in which a large 
number of simpler lattice models had been formulated earlier \cite{Gomsch} 
for the convenience of calculations. In this reformulation, a classical Ising 
spin variable $S$ is assigned to each lattice site; $S_i = 1$ ($-1$) if the 
$i$-th lattice site is occupied by a water (oil) molecule. If the $j$-th 
site is occupied by a monomer belonging to a surfactant then $S_j = 1, -1, 0$ 
depending on whether the monomer at the $j$th site belongs to head, tail or  
neutral part. The monomer-monomer interactions are taken into account through 
the interaction between the corresponding pair of Ising spins which is assumed 
to be non-zero provided the spins are located on the nearest-neighbour sites 
on the lattice. Thus, the Hamiltonian for the system is given by the 
standard form 
\begin{equation}
H = - J \sum_{<ij>} S_i S_j. 
\end{equation}
where attractive interaction (analogue of the ferromagnetic interaction in 
Ising magnets) corresponds to $J > 0$ and repulsive interaction (analogue 
of antiferromagnetic interaction) corresponds to $J < 0$ \cite{St1}. 
The temperature $T$ of the system is measured in the units of $J/k_B$  
where $k_B$ is the Boltzmann constant. 

Jan, Stauffer and collaborators \cite{St1} extended the model further to 
describe single-chain surfactants with ionic heads. According to their 
formulation, the monomers belonging to the ionic heads have Ising spin 
$+2$ to mimic the presence of electric charge. The repulsive interaction 
between a pair of ionic heads is taken into account through an 
(antiferromagnetic) interaction $J = -1$ between pairs of nearest neighbour 
sites both of which carry spins $+2$; however, the interaction between all 
other pairs of nearest-neighbour spins is assumed to be $J = 1$. By 
restricting the range of the repulsive (antiferromagnetic) interaction 
between the "charged" heads to only one lattice spacing one is, effectively, 
assuming very strong screening of the Coulomb repulsion between ionic heads 
by the counterions. 

Note that the monomers of the same surfactant as well as different 
surfactants are not allowed to occupy the same lattice site simultaneously; 
this represents a hard-core intra-chain as well as inter-chain repulsion 
for monomer separations smaller than one lattice spacing. Moreover, at 
any non-vanishing temperature, during the out-of-line thermal fluctuations 
of the chains, the hard-core repulsion leads to steric repulsion 
between the chains. Some interesting consequences of  steric 
repulsions between single-chain surfactants have been observed in 
earlier MC studies \cite{Chow1,Chow2,Chow3}. To our knowledge, no 
potential energies associated with the torsion of the surfactant 
chains have been incoporated so far in any work on Larson-type models.

\begin{figure}
\centerline{\psfig{file=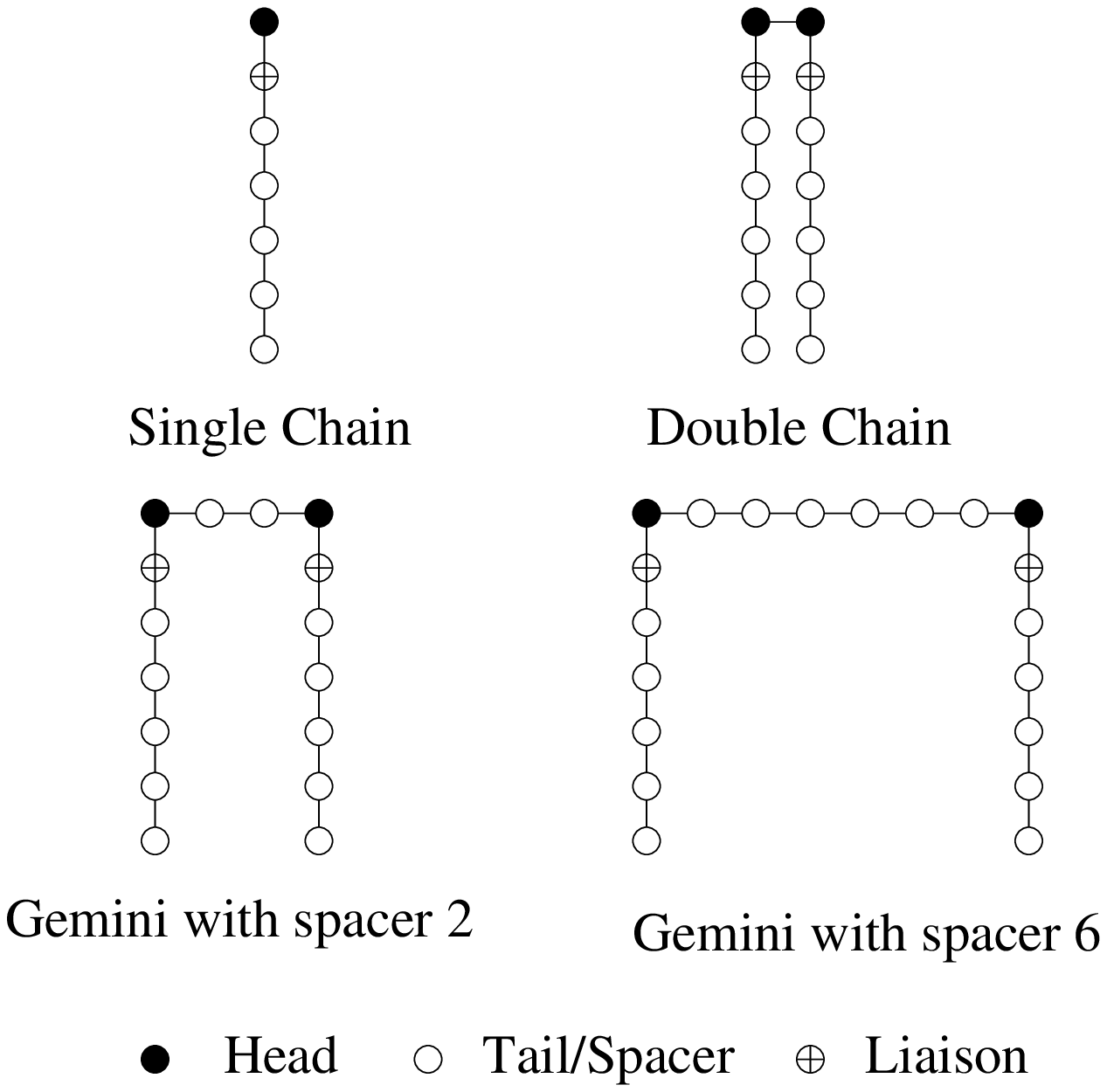,width=8cm}}
\caption{\sf Larson-type models of single-chain, double-chain
and gemini surfactants.}
\label{fig1b}
\end{figure}
Now we propose a microscopic lattice model of gemini surfactants.
In terms of the symbols used above to denote the primary "structure" of 
the microscopic lattice model of single-chain surfactants, Bernardes' 
lattice model of double-chain surfactants, with a single hydrophilic head, 
can be described by the symbol 
${\cal T}_m{\cal N}_p{\cal H}_q{\cal N}_p{\cal T}_m$. In terms 
of the same symbols, the microscopic lattice model of a gemini surfactant, 
which we propose here, can be represented by the symbol 
${\cal T}_m{\cal N}_p{\cal H}_q{\cal S}_n{\cal H}_q{\cal N}_p{\cal T}_m$ 
where $n$ is the number of lattice sites constituting the spacer represented 
by the symbol ${\cal S}$ (see fig.1(b)). 

Next, for the convenience of computation, we formulate the model in terms 
of classical Ising spin variables, generalizing the corresponding 
formulation for the single-chain surfactants reported in ref \cite{St1}. 
To our knowledge, all the gemini surfactants synthesized so far have  
ionic heads. Therefore, we incorporate the effects of the ionic heads 
following ref\cite{St1}; 
if the $j$-th site is occupied by a monomer belonging to a surfactant then 
$S_j = 1, -1, 0$ depending on whether the monomer at the $j$th site belongs 
to hydrophilic spacer, tail (or, hydrophobic spacer), neutral part, 
respectively while $S_j=+2$ if the $j$-th site is occupied by an ionic 
head. The monomer-monomer interactions are taken into account through 
the interaction between the corresponding pair of Ising spins 
the Hamiltonian for which is given by equation (1).

In order to predict the properties of gemini surfactants with non-ionic 
(polar) heads and to investigate which of the aggregation phenomena exhibited 
by the ionic gemini surfactants arise from the electric charge on their    
ionic heads, we have also considered a model of gemini surfactants with 
non-ionic polar heads which is obtained from the model of ionic gemini 
surfactants by replacing all the $+2$ Ising spin variables by Ising spin 
$+1$ (and, accordingly, the interactions $-1$ between the heads on 
nearest-neighbour sites are replaced by $+1$). Moreover, in order to 
investigate the role of the chain stiffness we have used a chain 
bending energy~\cite{Chow2}; every bend of a tail or a spacer, by a right 
angle at a lattice site, is assumed to cost an extra amount of energy 
$K (>0)$. 

Starting from an initial state (which will be described in the subsections 
2.1 and 2.2), the system is allowed to evolve following the standard 
Metropolis algorithm: each of the attempts to move a surfactant takes 
place certainly if $\Delta E < 0$ and with a probability proportional to 
$exp(-\Delta E/T)$ if $\Delta E \geq 0$, where $\Delta E$ is the change 
in energy that would be caused by the proposed move of the surfactant 
under consideration.

\begin{figure}
\centerline{\psfig{file=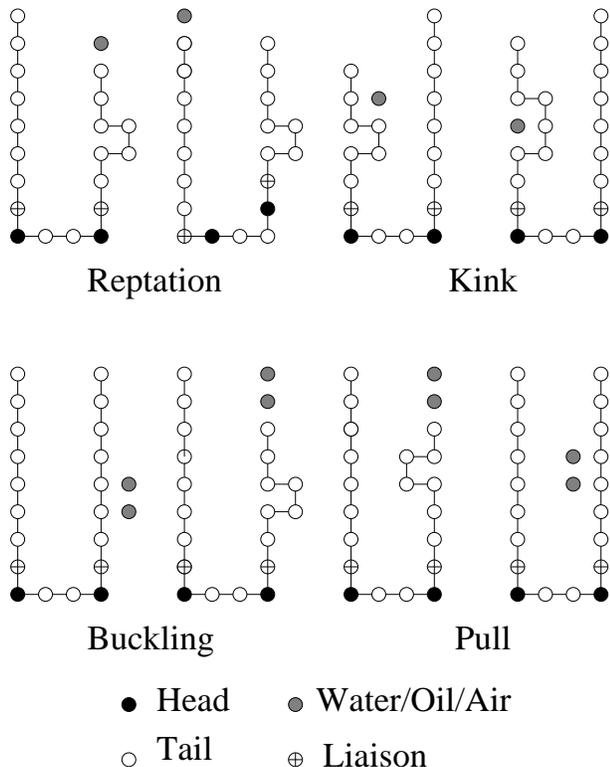,width=8cm}}
\caption{\sf Schematic representation of the moves attempted
by every surfactant at every MC step.}
\label{fig2}
\end{figure}
Next, we specify the allowed moves of the  surfactants for the appropriate 
sampling of the states of the system in a MC simulation. So far as the 
single-chain surfactants are concerned, the only move which was allowed in 
the pioneering works is\\ 
{\it reptation:} one of the two ends of each surfactant is picked up 
randomly, with equal probability, and the surfactant is allowed to move 
forward along its own contour by one lattice spacing with the probability 
mentioned above; this move effectively mimics the reptile-like slithering of 
the surfactants and hence the name. Subsequently, several new 
moves have been introduced in order to speed up the process of equilibration 
\cite{Livrev}. We have generalized these moves appropriately for the 
gemini surfactants and schematic representation of these moves are shown 
in fig.2; although the moves are illustrated using one of two tails, 
each of the moves (except reptation, which involves the entire surfactant) 
in our algorithm is equally likely to be applied on the two tails as well 
as the spacer. The moves allowed for the surfactants in our model are as 
follows:\\
(i) {\it reptation:} this is identical to the reptation move for single-chain 
surfactants described above; (ii) {\it spontaneous chain buckling:} a portion 
in the middle of one of the two tails or the spacer is randomly picked up 
and allowed to buckle with the probability mentioned above; (iii) {\it kink 
movement:} a kink formed by the buckling or reptation is allowed to move 
to a new position with the appropriate probability calculated according 
to the prescription mentioned above; (iv) {pull move:} this is the 
reverse of spontaneous chain buckling; a buckled part of one of the two 
tails or the spacer is pulled so as to make it more extended. In addition 
to these moves, each of the surfactants is allowed to move laterally 
in one of the six possible directions, which is chosen randomly with 
probability $1/6$, and each monomer of the surfactant is moved in that 
direction by one lattice spacing. each of these moves is possible only 
if the new positions of all the monomers are not occupied simultaneously 
by monomers belonging to other surfactants. Each surfactant is allowed 
to try each of the above mentioned moves once during each MC step.   
The moves of the molecules of water and air are described in 
the subsections 2.1 and 2.2 below.

In principle, one can study the aggregation of gemini surfactants deep 
inside bulk water and their spatial organization close to air-water 
interface by MC simulation of a single system where the lower part of 
the lattice representing water is sufficiently large and the density 
of surfactants is also sufficiently high so that a large fraction of 
these can be found deep inside water in the form of micellar aggregates. 
However, for the convenience of computation, we study these two aspects 
of the problem separately; in the first part we investigate only the 
phenomenon of aggregation of gemini surfactants in bulk water and in 
the second part we investigate only the spatial organization of the 
tails and spacers of the gemini surfactants at (and near) the air-water 
interface when the total volume fraction occupied by the surfactants 
is quite small. Both these parts of our computations are based on the 
general model described above and the specific differences involved in 
these two are explained separately in the next two subsections.    

\subsection{Model of Gemini Surfactants in Bulk Water:}

In order to investigate the spontaneous formation of micellar aggregates 
and their morphology, model gemini surfactants are initially dispersed 
randomly in a $L_x \times L_y \times L_z$ system which contains only water  
and surfactants. Periodic boundary condition is applied in all the three 
directions thereby mimicking bulk water which is infinite in all the 
directions. So far as the moves of the molecules of water are concerned, 
each molecule of water is allowed to exchange position with a monomer 
belonging to a surfactant, provided that is necessary for the implementation 
of an attempted move of a surfactant.

\subsection{Model of Gemini Surfactants at Air-Water Interface:}

Just as in the preceeding subsection, the system under investigation 
is modelled as a simple cubic lattice of size $L_x \times L_y \times L_z$. 
However, in contrast to simulating a bulk of water, which is infinite 
in all directions, we now simulate  a semi-infinite vertical column of 
air separated from a semi-infinite vertical column of water below by a sharp 
horizontal air-water interface. In the cartesian coordinate system we 
choose, the horizontal air-water interface is parallel to the $XY$-plane 
and the {\it vertically downward} direction is chosen as the $+Z$-axis. 
Each of the molecules of water and air can occupy a single lattice site. 
A classical Ising spin variable $S$ is assigned to each lattice site; 
$S_i = 1$ ($-1$) if the $i$-th lattice site is occupied by a water molecule 
(air molecule or empty). Our prescription for assigning the Ising spin 
variables to the sites occupied by the monomers of the amphiphiles is 
identical to that given in the preceeding subsection. Periodic boundary 
conditions are applied along the $X$ and $Y$ directions. The lattice sites 
in the uppermost and lowermost layers are occupied by "down" and "up" 
spins, respectively, which were not updated during the computer simulation. 
These boundary conditions mimic the physical situation, mentioned above, 
which we intend to simulate.  

In the initial state the surfactants are so arranged  that their 
spacers lie flat, and fully extended, horizontally in the first layer 
of water immediately below the air-water interface and their tails are 
fully extended vertically into air. The system is then allowed to 
evolve towards equilibrium following the Metropolis algorithm explained 
earlier. So far as the moves of the individual molecules are concerned, air 
and water are not allowed to exchange positions, as dispersion of air and 
water inside each other is not possible in our model. However, if some 
monomers of a surfactant come out of water the vacant sites are occupied 
by inserting water molecules; this is consistent with our assumption 
that the water column is semi-infinite in the $Z$-direction. Moreover,  
we impose an additional constraint that none of the heads of the surfactant 
molecules can come out of water. 

\subsection{Characteristic Quantities of Interest:} 

The most direct approach to investigate the morphology of the micellar 
aggregates and the spatial organization of the different parts of individual 
surfactants is to look at the snapshots of the system after equilibration. 
For studying the variations of CMC with the lengths of the tails and 
spacers one has to use a well-defined prescription for computing the CMC;
this is a subtle point as the CMC is not a unique single concentration, 
as mentioned before. We follow the prescription proposed, and used 
successfully in the case of single-chain surfactants \cite{St1}; we 
identify CMC as the amphiphile concentration where half 
of the surfactants are in the form of isolated chains and the other half 
in the form of clusters consisting of more than one neighbouring amphiphile.    

We have introduced a quantitative measure of the effective cross-sectional 
area $A$ of the gemini surfactants projected onto the air-water interface. 
We compute $|\Delta x|_m$ and $|\Delta y|_m$ which are the maximum 
differences in the $X$- and $Y$-coordinates, respectively, of the 
monomers and define $A$ as 
\begin{equation}
A = [(|\Delta x|_m)^2 + (|\Delta y|_m)^2]^{1/2} 
\end{equation}

The vertical extension $<Z>$ is defined as the difference in the $Z$-
coordinates of the highest and lowest monomers (i.e. monomers
with highest and lowest value of $Z$-coordinates) of a single 
surfactant ~\cite{Chow4}.
A quantitative measure of the gross features of the spatial organization 
of the tails and spacers of the gemini surfactants at the air-water 
interface is the equilibrium profiles of the concentrations of the 
corresponding monomers in the $Z$-direction, i.e., in the direction 
perpendicular to the air-water interface.  More precisely, at 
each molecular layer, we count separately the {\it number} of monomers 
belonging to the tails and the spacers (and also the heads and neutral 
parts) in that particular layer and average the data over sufficiently 
large number of configurations after equilibration of the system.

We have carried out MC simulations of the model 
${\cal T}_m{\cal N}_p{\cal H}_q{\cal S}_n{\cal H}_q{\cal N}_p{\cal T}_m$ 
of gemini surfactants for $p = q =1$ and for three different values of the 
tail length, namely, $m = 5, 15$ and $25$.   
In our simulation of the surfactants at the air-water interface we do not 
find any observable difference in the concentration profiles obtained in 
single runs for $100 \times 100 \times 100$ systems and for larger systems 
containing identical surface-density of surfactants, all the profiles 
reported in this paper have been generated for system sizes 
$L_x = L_y = L_z = L = 100$ by averaging over sufficiently large (10-25) 
number of runs. The same size of the system was also found to be large 
enough to avoid severe finite-size effects on the CMC data; each of the 
data points for the CMC is obtained by averaging over typically 10 runs.   
For a given $m$ we have computed the CMC for spacer lengths $2 \leq n \leq 20$. 

\section {Micellar Aggregates of Gemini Surfactants:}

\subsection {Aggregates of Gemini Surfactants: Results for Hydrophobic Spacers}
\begin{figure}
\centerline{\psfig{file=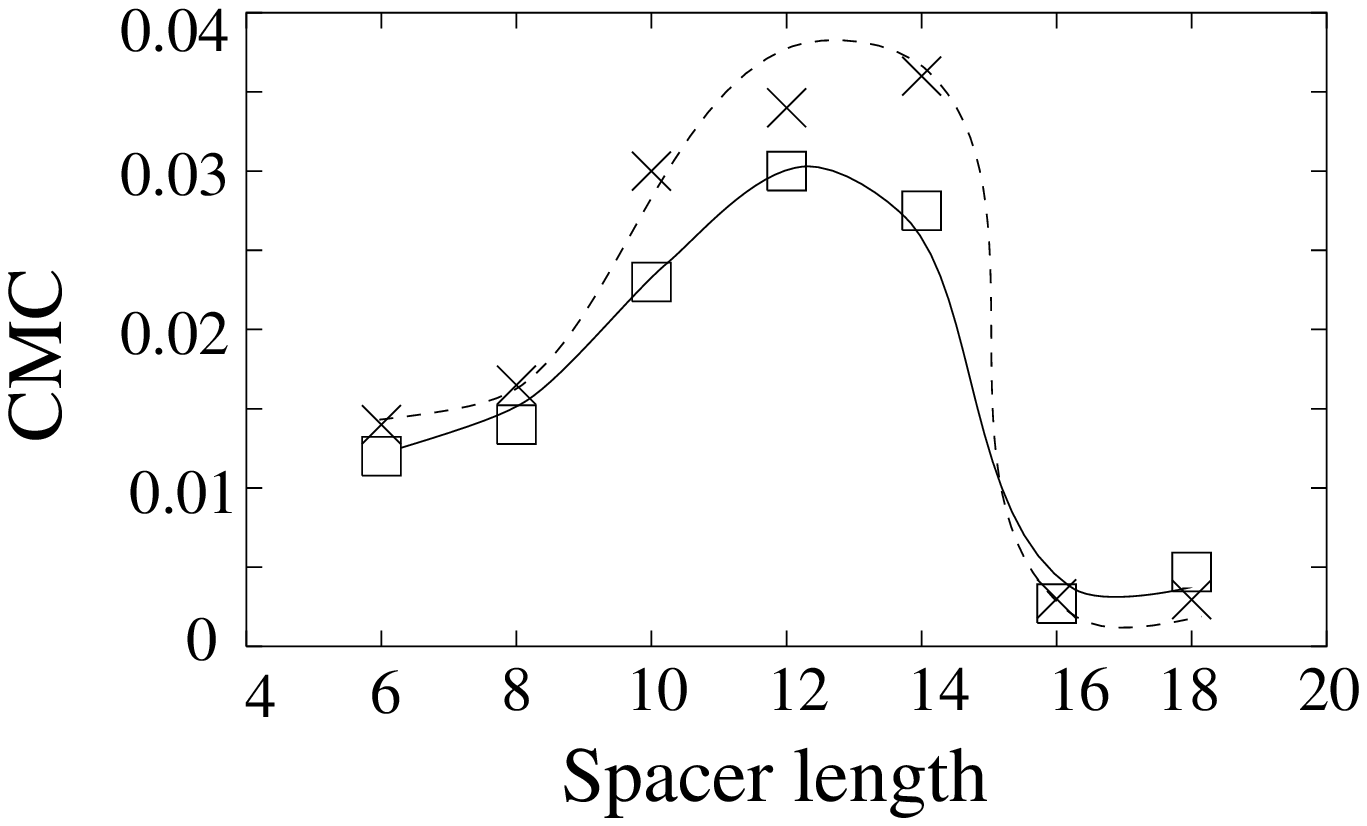,width=6cm,height=6cm}}
\centerline{\psfig{file=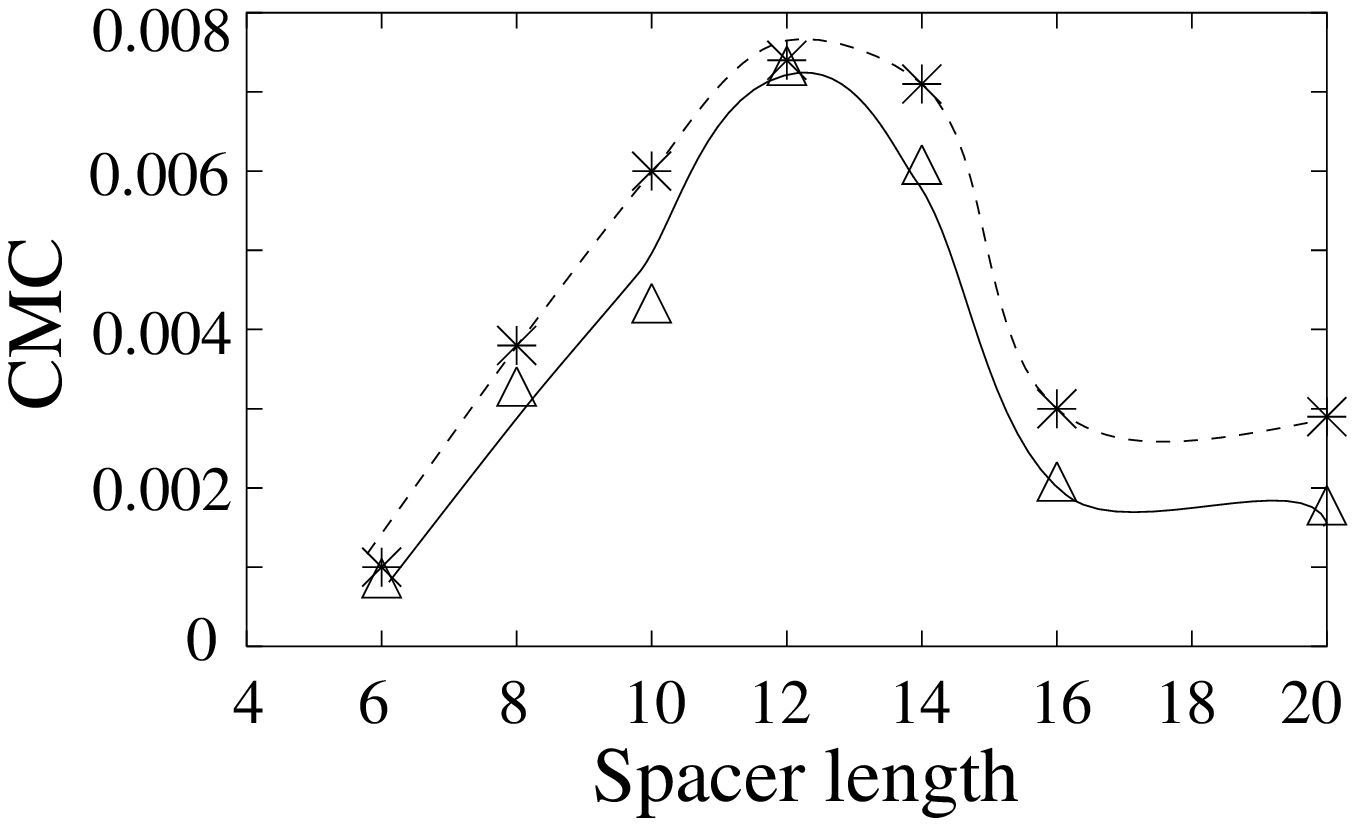,width=6cm,height=6cm}}
\caption{\sf (a)Variation of CMC of ionic geminis with hydrophobic
spacer length; $m = 15$, $T = 2.2$. The symbols $\Box$ and $\times$
correspond to $K = 0$ and $K = 2$, respectively.
(b) Same as (a), except that $m = 5$. The symbols
$\triangle$ and $\ast$ correspond to $K = 0$ and $K = 2$, respectively.
The continuous curves are merely guides to the eye.}
\label{fig3}
\end{figure}

The CMC of {\it ionic} gemini surfactants with {\it hydrophobic} spacers 
are plotted against the spacer length for two different lengths of the 
tail, namely, $m = 5$ and $m = 15$, in figs. 3 (a) and 3(b), respectively.  
The {\it non-monotonic} variation of CMC with the spacer length, in figs. 
3(a) and 3(b), is in qualitative agreement with the experimental observations 
\cite{Zana3,Frindi,goyel,Zana4}. However, this is in sharp contrast to 
the monotonic decrease of the CMC with the length of the hydrophobic 
tail of single-chain model surfactants of the type 
${\cal T}_m{\cal N}_p{\cal H}_q$ \cite{St1,Woer}. Moreover, for a given length of the 
hydrophobic spacer, the CMC of this type of gemini surfactants {\it 
increases} when the bending stiffness $K$ of the hydrophobic chains is 
switched on (see figs.  3(a) and (b)). Furthermore, we observe that, for 
a given length of the hydrophobic spacer, the CMC of ionic gemini 
surfactants {\sl increase} with the increase of the tail length (see fig.4); 
this trend of variation is also consistent with the corresponding 
experimental observations \cite{Menger1,Menger2}. 

\begin{figure}
\vspace*{1cm}
\centerline{\psfig{file=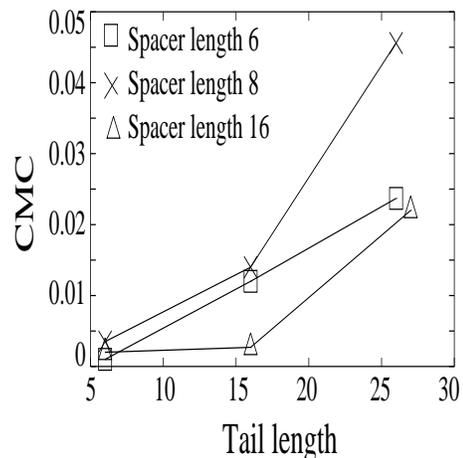,width=6cm,height=6cm}}
\caption{\sf  Variation of CMC of ionic geminis with tail
length at $T = 2.2$ for three different lengths of the hydrophobic
spacer, namely, $n = 6, 8, 16$. The straight lines connecting the
successive data points are merely guides to the eye.}
\label{fig4}
\end{figure}

\begin{figure}
\vspace*{1cm}
\centerline{\psfig{file=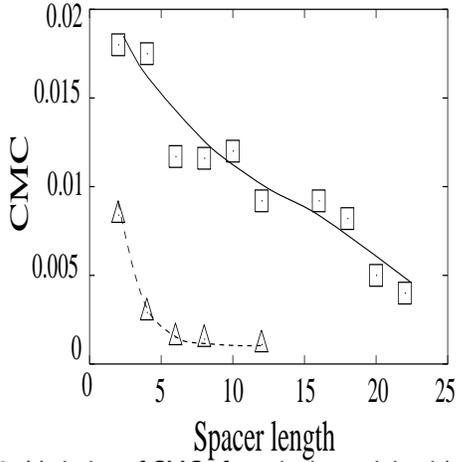,width=6cm,height=6cm}}
\caption{\sf Variation of CMC of non-ionic geminis with the
length of the hydrophobic spacer; $m = 15$ ($\Box$) and $m = 5$
($\triangle$) both with $K = 0$ and at $T = 2.2$. The continuous curves
are merely guides to the eye.}
\label{fig5}
\end{figure} 

For a given tail length (see fig.5 for $m = 5$ and $m = 15$), the CMC 
of model gemini surfactants with {\it hydrophobic} spacers decreases 
{\it monotonically} with the increase in the spacer length when the polar 
head group is {\it non-ionic}. This is in sharp contrast to the 
non-monotonic variation observed for ionic gemini surfactants. However, 
for a given spacer length, the trend of the variation of CMC of non-ionic 
gemini surfactants with the tail length is similar to that observed for ionic 
gemini surfactants, i.e., CMC increases with the increase of the length 
of the tail. 

\begin{figure}
\centerline{\psfig{file=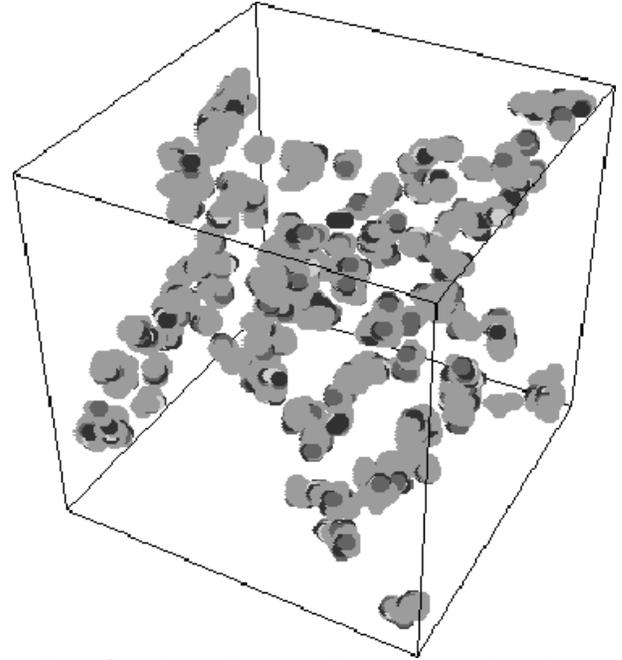,width=8cm}}
\caption{\sf  Snapshots of the micellar aggregates formed by
ionic geminis with hydrophobic spacer; $m = 15$, $n = 2$ and $K = 0$ at
$T = 2.2$ when  the surfactant density is $0.007$.
The symbols black spheres, dark grey spheres and light grey spheres
represent monomers belonging to head, tail and spacer, respectively.}
\label{fig6}
\end{figure}

\begin{figure}
\centerline{\psfig{file=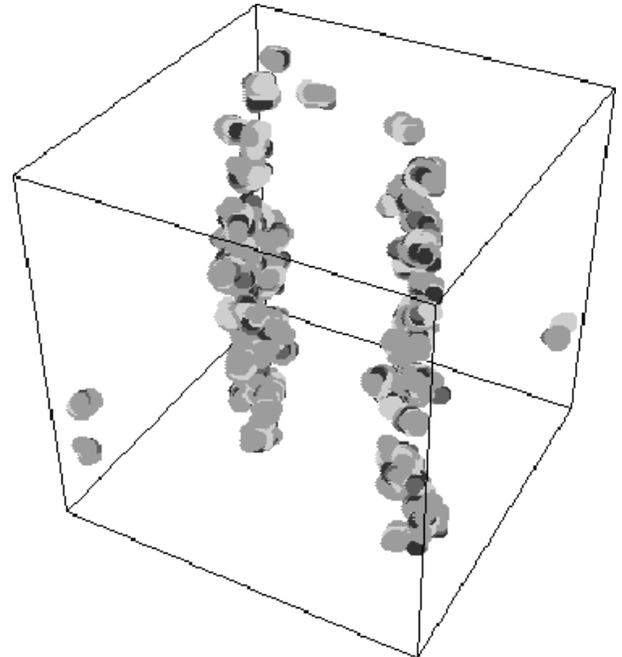,width=8cm}}
\caption{\sf Same as in fig. 6, except that $n = 16$ and the
density is $0.005$.}
\label{fig7}
\end{figure}

Instantaneous snapshots of the micellar aggregates formed by {\it long 
tailed} ($m = 15$) gemini surfactants with {\it ionic heads} and 
{\it hydrophobic spacer} are shown for spacer lengths $n = 2$ (fig. 6) 
and $n = 16$ (fig.7). The morphology of the aggregates in fig.6 are 
similar to the "long, thread-like and entangled" micelles observed in 
laboratory experiments \cite{Zana2} and in MD simulations \cite{Karaborni} 
on gemini surfactants with short hydrophobic spacers. Moreover, our data 
in fig.7 suggest that rod-like ("columnar") micelles are formed by gemini 
surfactants with {\it ionic head} and {\it long tail} ($m=15$) when the 
length of the {\it hydrophobic spacer} is also long ($n = 16$). The 
morphologies of the aggregates in fig.6 and 7 are in sharp contrast with 
the spherical shape of the micelles (see fig.8) formed by single-chain 
ionic surfactants of comparable tail size even at concentrations 
somewhat higher than those in the figures 6 and 7.

\begin{figure}
\centerline{\psfig{file=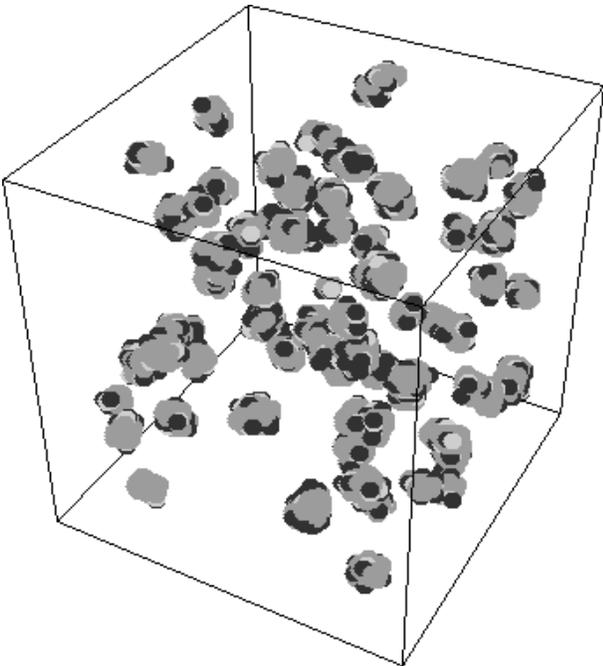,width=8cm}}
\caption{\sf Snapshots of micellar aggregates formed by
single-chain ionic surfactants with $m = 14$ and the density $0.01$.
The symbols black spheres and grey spheres represent monomers
belonging to head and tail, respectively.}
\label{fig8}
\end{figure}

\begin{figure}
\centerline{\psfig{file=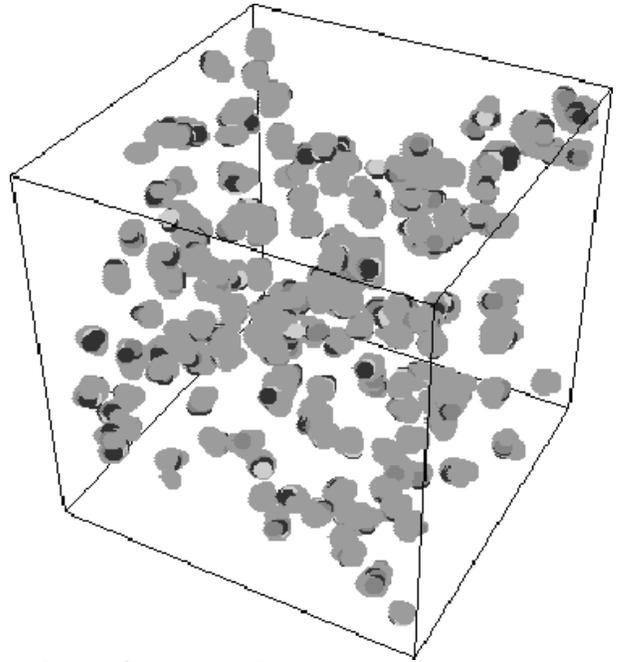,width=8cm}}
\caption{\sf Same as in fig. 6, except that the geminis
are non-ionic.}
\label{fig9}
\end{figure}
There is no significant difference between the morphologies of the  
micellar aggregates of ionic and non-ionic single-chain model surfactants 
represented by the symbol ${\cal T}_m{\cal N}_p{\cal H}_q$ \cite{Maiti}. 
Similarly, we do not observe any significant difference also in the shapes 
of the aggregates of ionic surfactants (figs.6 and 7) and those of 
non-ionic gemini surfactants (see figs.9 and 10) with hydrophobic spacers, 
for given values of $m$, $n$ and comparable concentration, in spite of 
qualitatively different trends of variation of their CMCs with spacer 
lengths. 

\begin{figure}
\centerline{\psfig{file=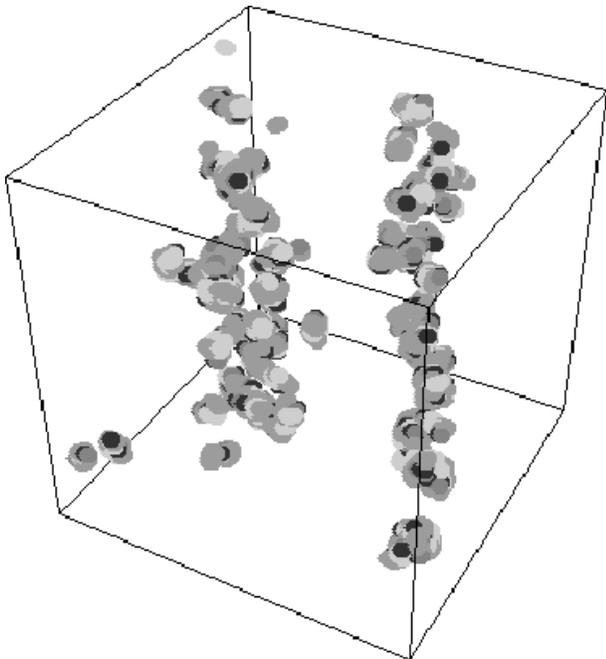,width=8cm}}
\caption{\sf Same as in fig. 7, except that the geminis
are non-ionic.}
\label{fig10}
\end{figure}

\begin{figure}
\centerline{\psfig{file=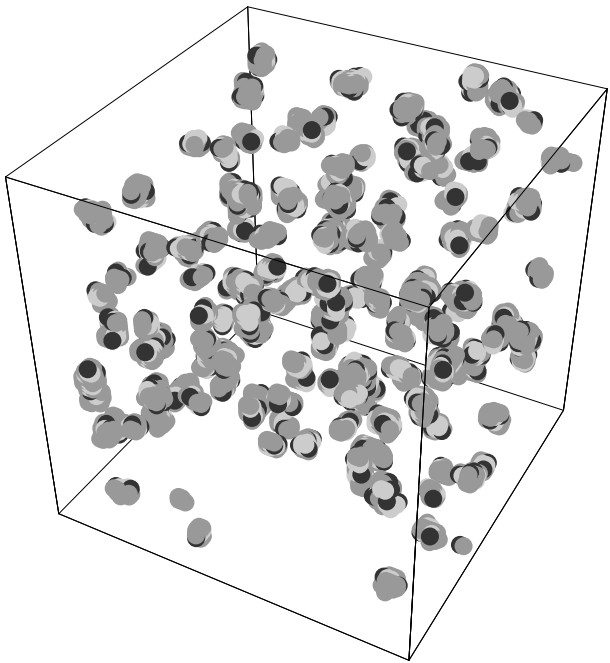,width=8cm}}
\caption{\sf Snapshots of micellar aggregates formed by
ionic gemini surfactants with hydrophobic spacer; $m = 5$, $n = 6$ and
$K = 0$ at $T = 2.2$ when the surfactant density is $0.005$.}
\label{fig11}
\end{figure}
Note that the rod-like micelles, shown in fig.7, are formed when the 
length of the hydrophobic spacer and the combined length of the 
tail and the neutral part of the gemini surfactants are both equal to 
$16$. Does this imply that rod-like micelles are formed whenever the 
hydrophobic spacer and the tail are equal (or comparable) in length? 
In order to answer this question we have also looked at the snapshots 
of the micellar aggregates of similar gemini surfactants with shorter 
tails and spacers; a typical example, shown in fig.11, corresponds 
to $m = 5$, $n = 6$. The fact that these micelles are also "long, 
thread-like and entangled", like those in fig.6, in contrast to the 
rod-like micelles of fig.7,  suggests that the morphology of the ionic 
gemini surfactants with hydrophobic spacers is dominantly determined by 
the length of the spacer; long, thread-like micelles are formed if 
the spacer is short and rod-like micelles are formed if the spacer is 
long.  

\subsection {Aggregates of Gemini Surfactants; Results for Hydrophilic Spacers}
\begin{figure}
\vspace*{1cm}
\centerline{\psfig{file=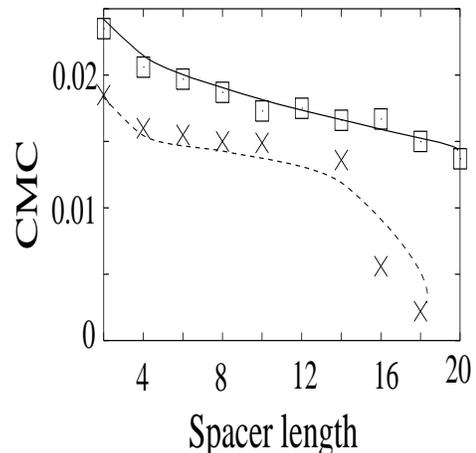,width=6cm,height=6cm}}
\caption{\sf (a)Variation of CMC of ionic geminis with hydrophilic
spacer length; $m = 15$, $T = 2.2$. The symbols $\Box$ and $\times$
correspond to $K = 0$ and $K = 2$, respectively.}
\label{fig12}
\end{figure}

In fig.12 we plot the CMC  against the length of the {\it hydrophilic} 
spacer for gemini surfactants with {\it ionic} head and tail length 
$m = 15$ (the qualitative features of the corresponding curve for $m = 5$ 
are very similar and, therefore, not shown). In contrast to the non-monotonic 
variation of CMC observed earlier with the variation of the length of 
hydrophobic spacers, now we find a {\it monotonic} decrease of CMC with 
the increase of the length of the hydrophilic spacer.  This trend of 
variation is in qualitative agreement with the corresponding 
experimental observation \cite{Zhu}. Moreover, for given lengths of the 
hydrophobic spacer and the tail, $n$ and $m$, the CMC for the bending 
energy $K = 2$ is lower than that for $K = 0$ (see fig.12); this trend 
of variation is exactly opposite to the corresponding trend observed 
earlier in the case of gemini surfactants with hydrophobic spacers. 

\begin{figure}
\centerline{\psfig{file=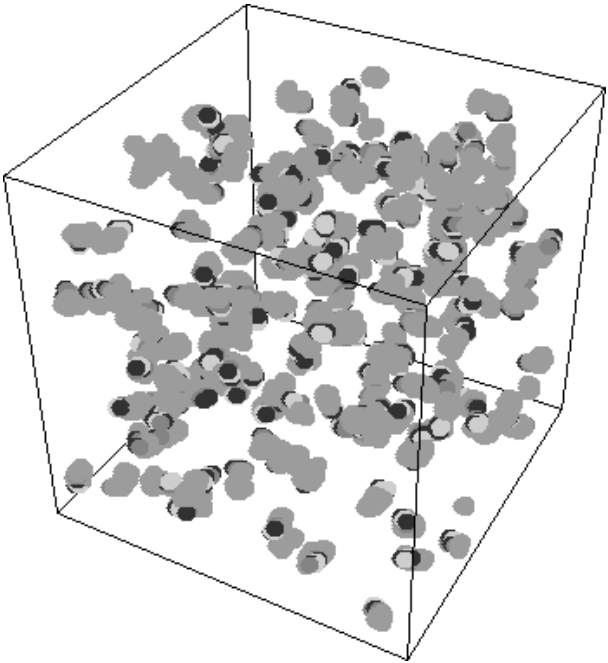,width=8cm}}
\caption{\sf Same as fig. 6 except that the spacer is hydrophilic.}
\label{fig13}
\end{figure}

\begin{figure}
\centerline{\psfig{file=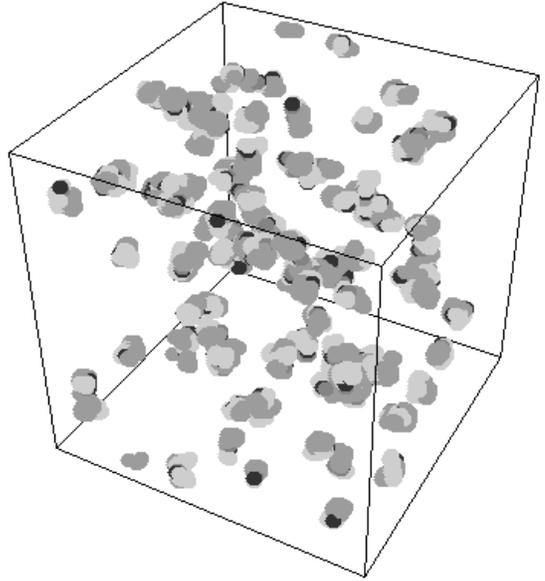,width=8cm}}
\caption{\sf Same as fig. 7 except that the spacer is hydrophilic.}
\label{fig14}
\end{figure}

The snapshots of the micellar aggregates formed by the gemini surfactants 
with {\it ionic heads} and {\it hydrophilic spacer} are shown for spacer 
lengths $n = 2$ (fig. 13) and for $n = 16$ (fig.14) for densities which 
are identical to those in the figs.6 and 7, respectively. Comparing the 
morphologies of the aggregates in fig.6 and fig.13 we find that the 
gemini surfactants with hydrophobic spacers form coarser (albeit fewer 
in number) aggregates compared to the corresponding geminis with 
hydrophilic spacers; this is also consistent with the fact that the CMC 
of the gemini surfactants with spacer length $n = 2$ are higher when the 
spacers are hydrophilic as compared to that for hydrophobic spacers.  

The difference in the morphologies of ionic geminis with hydrophobic and 
hydrophilic spacers is much more striking when the spacer is longer 
($n = 16$)(compare the fig.7 with fig.14)- the micelles are more or less 
spherical when the spacers are hydrophilic! 

An important difference between the micellar aggregates of gemini 
surfactants with hydrophobic spacers and those with hydrophilic spacers 
is that more spacer monomers are found on the outer surface of the 
aggregate (i.e., in contact with water) when the spacer is hydrophilic. 
This is consistent with one's intuitive expectation because the 
hydrophilic spacers like to be in contact with water.  

The snapshots of the micellar aggregates of non-ionic gemini surfactants 
with hydrophilic spacers are very similar to those for the corresponding 
ionic gemini surfactants (and, therefore, not shown in any figure).

Hydrophilic spacers gain energy by remaining surrounded by water.
On the other hand, hydrophobic spacers as well as tails try to avoid contact
with water by hiding inside micellar aggregates. That is why
in the snapshots of micellar aggregates we see that a larger number
of monomers belonging to the spacers are in contact with water, when
the spacers are hydrophilic, than those when the spacers are hydrophobic.
And this is prominent particularly for long spacers.

\section {Spatial Organization of Gemini Surfactants at Air-Water Interface: 
Results for Hydrophobic and Hydrophilic Spacers}

\subsection{Dilute regime} 

\begin{figure}
\vspace*{1cm}
\centerline{\psfig{file=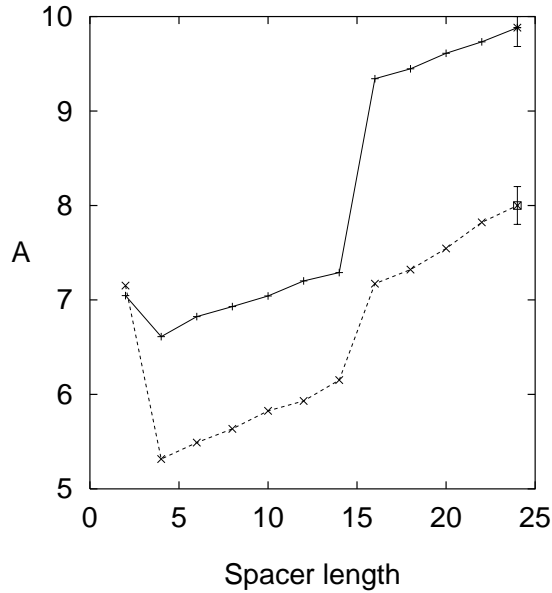,width=8cm}}
\caption{\sf Variation of cross-sectional area of isolated individual gemini
surfactant with spacer spacer length. The solid line for hydrophobic
spacer and broken line for hydrophilic spacer.
To give an indication of the accuracy of the data points the error bar
of only one point has been shown.}
\label{fig15}
\end{figure}

First let us consider the dilute regime where the concentration of 
the surfactants is so low that not only all of them remain, almost 
certainly, at the air-water interface but every surfactant may be 
regarded as, effectively, isolated from each other. In this limit the 
cross-sectional area $A$ of the molecules is determined by only 
intra-molecular interactions, which is dominated by the steric 
(entropic) interactions among the tails and the spacer.  We plot the 
cross-sectional area $A$ of isolated individual gemini surfactants as 
a function of the length of the spacer in fig.15, in both the cases of 
(a) hydrophobic and (b) hydrophilic spacers. 
The spacer is very stiff when its length is $n = 2$ as no wiggle can form. 
The area $A$ for $n = 4$ is smaller than that for $n = 2$ irrespective of 
the nature of the spacer (i.e., hydrophobic or hydrophilic); this is 
caused by the formation of wiggle on the spacer which brings the two 
heads closer. Further increase of the spacer length gives rise to a 
linear increase of the area $A$. However, a sharper increase in $A$ 
takes place when the length of the spacer becomes equal to that of the 
tails; on both sides of this regime of sharp rise, the rate of increase 
of $A$ with $n$ is practically identical. 

Because of its stiffness against wiggle formation, the spacer of 
length $n = 2$ can buckle neither towards air nor towards water 
and remains parallel (like a rigid rod) to the air-water interface. 
Therefore, if $n = 2$, the cross-sectional area $A$ of isolated 
gemini surfactants with hydrophilic spacers is practically identical  
to that of gemini surfactants with hydrophobic spacers.
However, for all larger values of $n$, $A$ is smaller if the spacer 
is hydrophilic; a hydrophilic spacer buckles into water thereby 
leaving most of the space in the air above the heads available for 
occupation by the tails. On the other hand, the hydrophobic tails 
take up a substantial amout of available space in a cap-like volume 
in the air just above the heads thereby forcing the tails to spread 
out radially outward and, hence, increasing the effective area $A$. 

\begin{figure}
\vspace*{1.5cm}
\centerline{\psfig{file=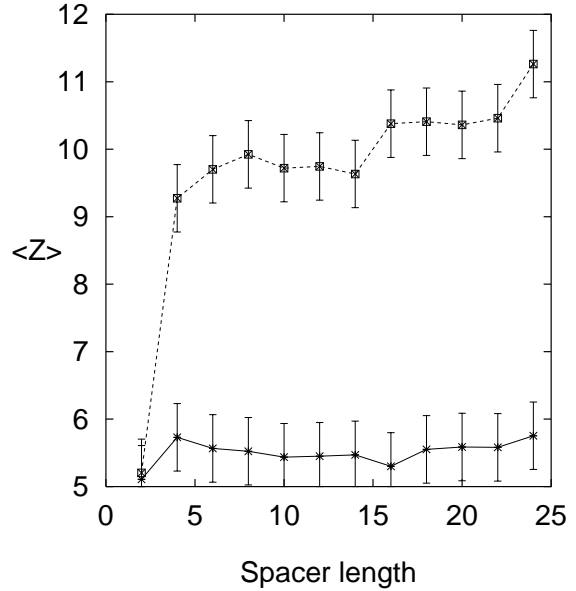,width=8cm}}
\caption{\sf Variation of vertical extension of individual gemini
surfactant with spacer length. The solid line for hydrophobic
spacer and broken line for hydrophilic spacer.}
\label{fig16}
\end{figure}
Evidence in support of this scenario emerges also from the plots of 
vertical extension $V$ of the isolated gemini surfactants against 
the length of the spacer (see fig.16); a larger $V$ of gemini 
surfactants with hydrophilic spacer, as compared to those of gemini 
surfactants with hydrophobic spacer of identical length, arises from 
the fact that the hydrophilic spacers buckle into water while their 
tails remain outside water.     

\subsection{High Surface Density regime}

\begin{figure}
\vspace*{1cm}
\centerline{\psfig{file=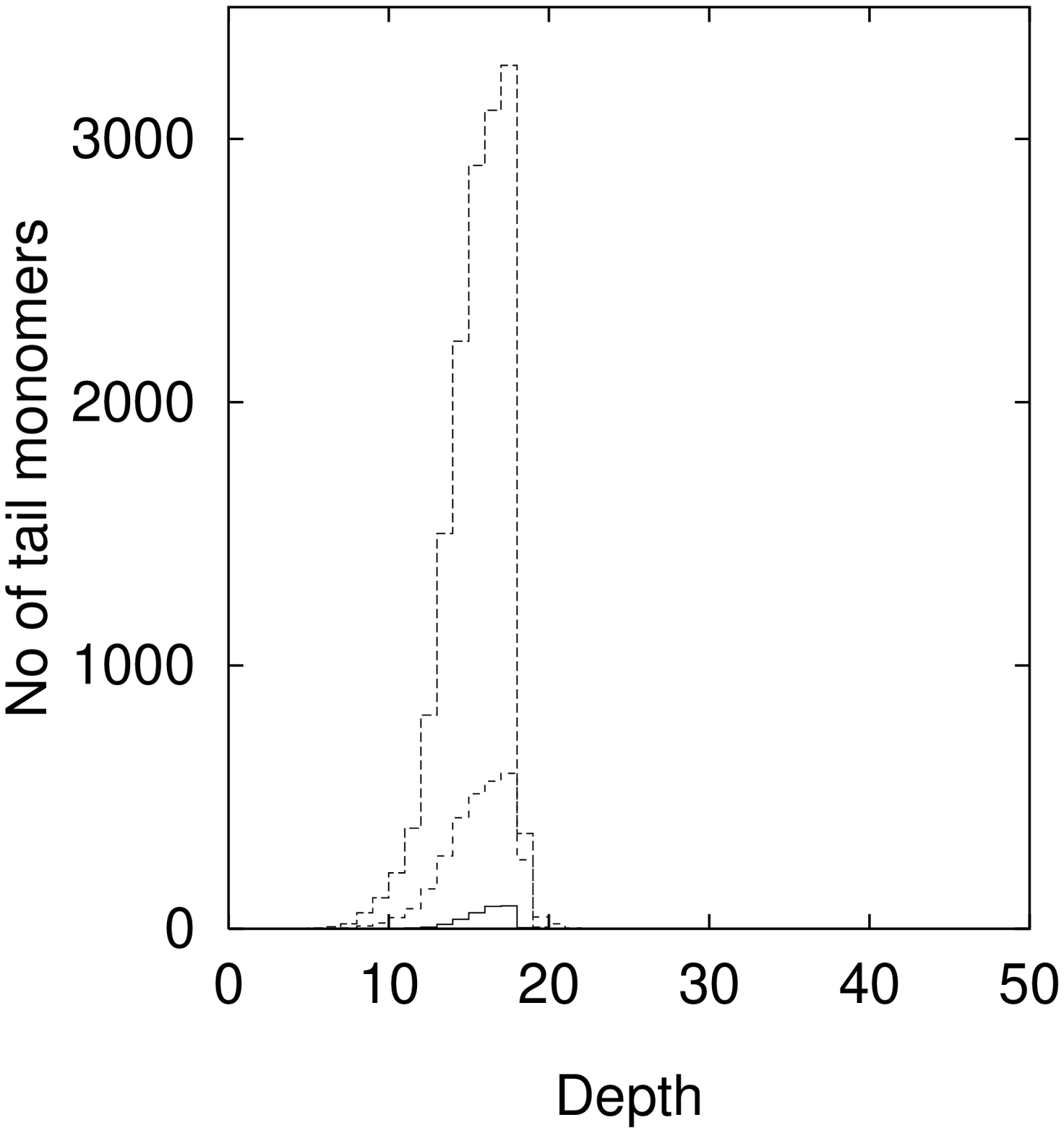,width=8cm}}
\centerline{\psfig{file=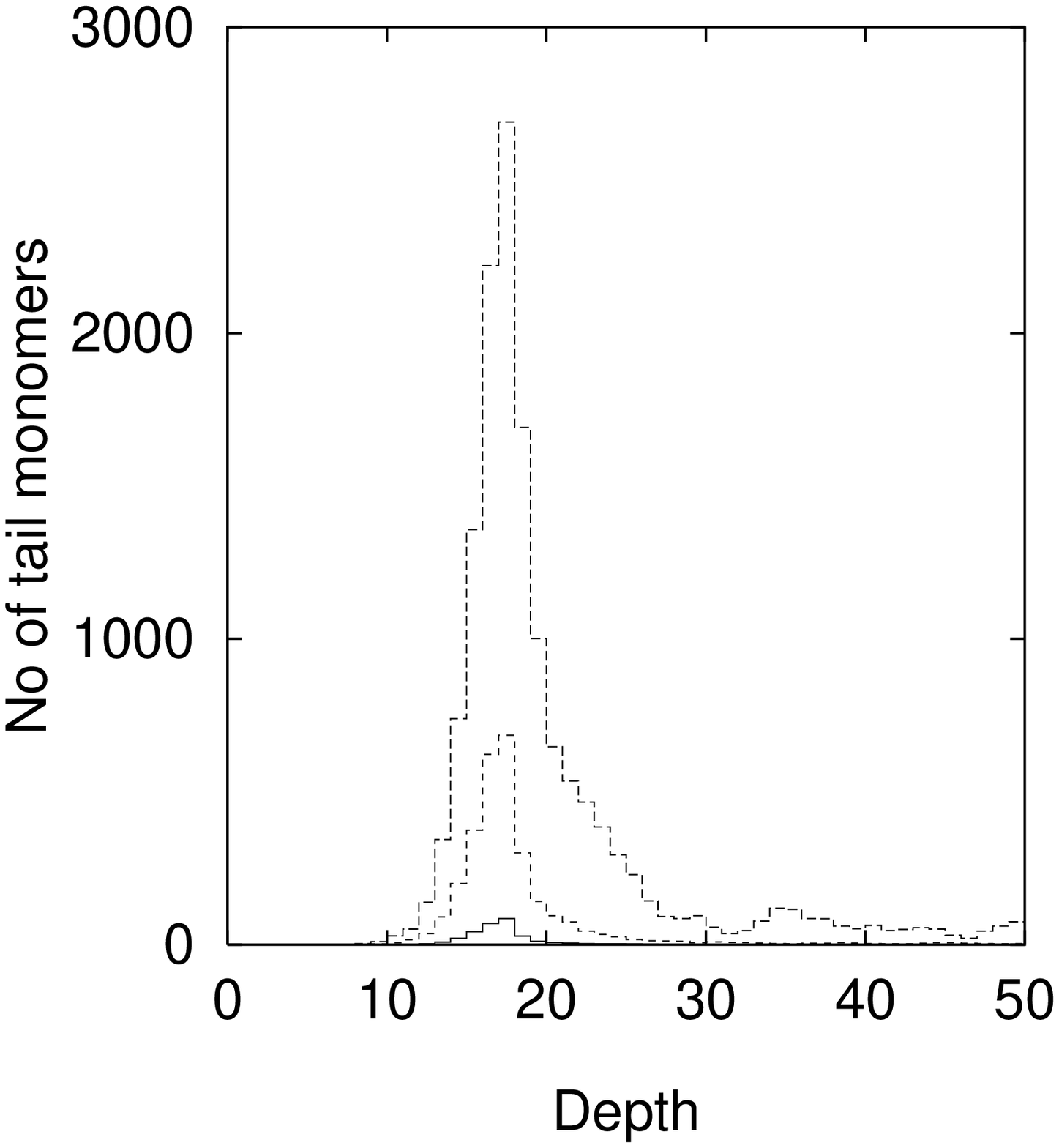,width=8cm}}
\caption{\sf Concentration profiles for tail monomers for three
different cases when the number of amphiphiles present in the systems
are (i) $500$ (solid line), (ii) $100$ (broken line) and (iii)
$10$ (dashed line).(a) hydrophobic spacer (b) hydrophilic spacer.}
\label{fig17}
\end{figure}

The concentration profiles of the tails of the gemini surfactants with 
hydrophobic spacers are shown in fig.17(a) and  the corresponding 
concentration profiles for gemini surfactants with hydrophilic spacers 
are shown in fig.17(b). In the case of gemini surfactants with hydrophobic 
spacer, the spacers minimize contact with water by arranging themselves 
just outside the water, but do not venture out too far from the interface. 
On the other hand, the hydrophilic spacers gain energy by moving inside 
water thereby leaving more space just outside water which become 
available for occupation by the tails; consequently, one would have 
naively expected, the tails of the gemini surfactants with hydrophilic 
spacers are likely to be found closer to the interface that those of 
the gemini surfactants with hydrophobic spacers. However, what we 
observe in reality in fig.17(b) is much more dramatic- a significantly 
large fraction of the monomers belonging to the tails are pulled into water 
along with the heads (see fig.18) to which they are attached! The loss of 
energy due to the increase in the area of contact between the hydrophobic 
tails and warter is compensated by the gain of energy from the increase 
of contact between hydrophilic spacers and water as well as the gain of  
conformational  entropy of the system arising from the larger amount of space 
available to those chains which remain at the interface. This interpretation 
is supported by our observation that this effect  is more prominent at 
higher densities of surfactants.
Some other manifestations of the entropic effect have been observed 
earlier\cite{Chow1,Chow2,Chow3}. 

\begin{figure}
\vspace*{1.5cm}
\centerline{\psfig{file=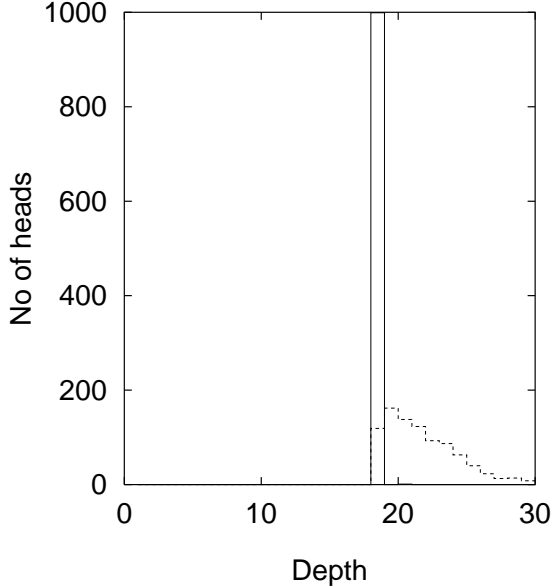,width=8cm}}
\caption{\sf Concentration profiles for heads when the number of amphiphiles
is $500$. Solid line is for hydrophobic spacer and broken line for
hydrophilic spacer.}
\label{fig18}
\end{figure}
The conclusions drawn from averaged concentration profiles are supported 
by the instantaneous snapshots of the surfactants (not shown in any 
figure).

\section{Summary and Conclusion:} 

In this paper we have developed models of both ionic and non-ionic 
gemini surfactants with hydrophobic spacers as well as those with 
hydrophilic spacers. We have investigated the morphologies of the 
micellar aggregates of these gemini surfactants and computed the 
corresponding CMCs by carrying out MC simulations. 

The main features of the aggregation of gemini surfactants with 
{\it hydrophobic} spacers can be summarized as follows: 
(i) the micelles are far from spherical- for short spacers these are long 
"thread-like" and for long spacers these are "rod-like"; 
(ii) the CMC varies non-monotonically with increasing spacer length; 
(iii) the CMC increases with the increase of the bending stiffness of the 
tails and spacers.  

The main features of the aggregation of gemini surfactants with 
{\it hydrophilic} spacers can be summarized as follows: 
(i) the micelles are more or less spherical; (ii) the CMC decreases 
monotonically with increasing spacer length; (iii) the CMC decreases with 
the increase of the bending stiffness of the tails and spacers.  

In contrast to the case of single chain surfactants the CMC increases
with the hydrocarbon tail length for both the ionic and non-ionic
gemini surfactants irrespective of whether the spacer is hydrophobic
or hydrophilic. However like the case of single chain surfactants
the morphologies
of the ionic gemini surfactants are identical to that of the corresponding
non-ionic gemini surfactants both for hydrophobic as well as hydrophilic
spacer.

Therefore, we conclude that (i) the shapes of aggregates are dominantly 
determined by the geometric shape and size of the molecules and 
whether the spacer is hydrophobic or hydrophilic, whereas 
(ii) the variation of CMC with spacer length is strongly influenced by the 
ionic charge and, again, whether the spacer is hydrophobic or hydrophilic. 

In the case gemini surfactants at the air-water interface
for dilute regime, the cross-sectional area for single isolated gemini
surfactant increases with the spacer length both for hydrophobic and hydrophilic
spacer. However beyond a certain length of the spacer the cross-sectional
area is larger for the hydrophobic spacer as compared to that for
hydrophilic spacer. These trends are consistent with the variation
of vertical extension $<Z>$ with spacer length; a larger value of $<Z>$ for
hydrophilic spacer as compared to the hydrophobic spacer of identical
length is observed in our simulations.

For extremely high surface density of surfactants at the air-water
interface we have demonstrated qualitatively the spatial organization
of the gemini surfactants for both the case of hydrophobic and
hydrophilic spacers.
 
In view of the above observations, it seems that the main effects
of introducing the spacer is to impose an additional geometrical
constraint on the packing of surfactant molecules and, therefore,
to influence their aggregate shape and other properties.

Molecular dynamics (MD) simulations of a similar molecular model of gemini 
surfactants has been carried out by Karaborni et al. \cite{Karaborni}. 
In their model, particles of water interact mutually via a truncated 
Lennard-Jones (LJ) potential with sufficiently long cut-off to incorporate 
both the short-range repulsion and long-range attraction. The mutual 
interactions between the pairs of particles belonging to the tail were also 
simular. But, the cut-off range of the tail-water and head-head interactions 
were so short that no attraction was possible. However, the chains and 
spacers simulated by Karaborni et al. were much smaller than those 
investigated in our paper here. Besides, Karaborni et al. neither  
investigated the CMC and its variations with lengths of the tails and 
spacer nor considered any model of gemini surfactants with hydrophilic 
spacers. One should also try to develope more efficient MD algorithms to 
repeat our computations with more realistic interaction potentials 
on a continuum to check if any of the morphologies observed in this 
paper have been influenced significantly by the discrete lattice.

It would be interesting to investigate the effects of 
weakening of the screening (i.e., increasing the range) of the repulsive 
Coulomb interaction between the ionic heads on the results reported in this 
letter; but, such a MC study will require much larger computational 
resources.

{\bf Acknowledgements:} One of us (DC) thanks V.K. Aswal, A.T. Bernardes, 
S. Bhattacharya, P.S. Goyal, D. Stauffer and R. Zana for enlightening  
discussions/correspondences and the Alexander von Humboldt Foundation for 
supporting the computations, carried out at IITK on the gemini surfactants 
in bulk water, through a research equipment grant. DC also thanks D. Stauffer 
for the hospitality, the Humboldt Foundation and SFB K\"oln-Aachen-J\"ulich 
for financial support during a stay in Cologne where a large part of the 
computations were carried out. We also thank D. Stauffer for a critical 
reading of an earlier version of the manuscript, for comments and 
suggestions, and 
P. Guptabhaya for allowing us to share his computer graphics packages for 
producing some of the snapshots of the aggregates.

\end{document}